\documentclass[ ,final            
  ]
  {aipproc}
\layoutstyle{6x9}
\begin{document}
\title{Spin Sum Rules and Polarizabilities}
\author{D. Drechsel}
{address={Institut für Kernphysik, Universität Mainz, 55099 Mainz, Germany}}
\begin{abstract}
The Gerasimov-Drell-Hearn sum rule and related dispersive integrals
connect real and virtual Compton scattering to inclusive photo- and electroproduction.
Being based on universal principles as causality, unitarity, and gauge invariance, these relations
provide a unique testing ground to study the internal degrees of freedom that
hold a system together. The present contribution reviews the
spin-dependent sum rules and cross sections of the nucleon. At small
momentum transfer, the data sample information on the long range
phenomena (Goldstone bosons and collective resonances), whereas the primary degrees of freedom
(quarks and gluons) become visible at large momentum transfer (short distance). The rich
body of new data covers a wide range of phenomena from coherent to incoherent
processes, and from the generalized spin polarizabilities on the low-energy
side to higher twist effects in deep inelastic scattering.
\end{abstract}
\maketitle
\section{Introduction}
The spin structure of the nucleon has been at the forefront of hadronic physics
ever since Stern and collaborators~\cite{Frisch:1933dd} discovered the large anomalous moment
of the proton. This pilot experiment indicated that nucleons are composite systems
with internal degrees of freedom, leading also to a
finite size and a rich excitation spectrum of protons and neutrons. Since the 1950s, these aspects
have been under active experimental investigation with the electromagnetic probe.
Modern electron accelerators combined with new polarized beam and target
techniques have provided a wealth of new precision data. Because of the polarization degrees of freedom,
these data form a solid basis for a detailed picture of the nucleon's spin structure.
For real photons, the GDH Collaboration at MAMI and ELSA confirmed the Gerasimov-Drell-Hearn (GDH)
sum rule~\cite{Gerasimov:1965et,Drell:1966jv} and determined the forward spin polarizability (FSP) of the proton,
which is an important input for real Compton scattering (RCS). Collaborations in all three halls of
Jefferson Lab collected a rich body of precision data
for electroproduction with polarized beams and targets. As a result, the evolution of generalized
GDH-like integrals and polarizabilities is now known over a large range of 4-momentum transfer $Q^2$. These
recent developments have
bridged the gap between the high-$Q^2$ (short range) aspects of deeply inelastic scattering (DIS) on
the nucleon's constituents and the low-$Q^2$ (long range) phenomena like resonance excitation. In
the present contribution to the ``Spin Structure at Long Distance'' workshop, the focus is on
long-range aspects of the nucleon's spin structure.
For detailed information on this field, the reader is referred to the following review papers:
an introduction to doubly-virtual Compton scattering (VVCS) by Ji and Osborne~\cite{Ji:1999mr},
a review on dispersion theory in RCS and VVCS~\cite{Drechsel:2002ar},
a report on the GDH sum rule and related integrals~\cite{Drechsel:2004ki},
and the recent review by Kuhn, Chen, and Leader~\cite{Kuhn:2008sy}.
\section{Forward RCS and Photoproduction}
\subsection{RCS formalism}
The incident photon is described by its 4-momenta of momentum, $q=(|{\vec{q}}|,\,{\vec{q}})$, and polarization,
$\varepsilon_{\lambda}=(0,\,\vec {\varepsilon}_{\lambda})$, which obey the relations $q\cdot q=0$ (real photon),
$\varepsilon_{\lambda}\cdot q=0$ (transverse gauge), and $\vec {\varepsilon}_{\lambda}^{\ast} \cdot \vec {\varepsilon}_{\lambda}=-1$.
If the incoming photon moves in the direction of the $z$-axis,
a circularly polarized photon is characterized by helicities $\lambda=+1$ (right-handed) and $\lambda=-1$ (left-handed),
and ${\vec {\varepsilon}}_{\pm} = \mp ({\hat {e}}_x \pm i{\hat {e}}_y)/\sqrt{2}$. Furthermore, the incident photon energy in the
lab frame is denoted by $\nu$ and the outgoing photon is described by the polarization vector $\vec {\varepsilon}_{\lambda}'$.
The absorption of the photon leads to an excited state with total c.m. energy $W=\sqrt{M^2+2M\nu}$,
where $M$ is the nucleon mass. The forward Compton tensor takes the general form
\begin{equation}
\label{eq1.1}
T(\nu) = {\vec {\varepsilon}}'^{\ast}\cdot{\vec {\varepsilon}}\; f(\nu)+
i\,{\vec {\sigma}}\cdot({\vec {\varepsilon}}'^{\ast}\times {\vec {\varepsilon}})\; g(\nu)\, .
\end{equation}
The Compton tensor is invariant under crossing , $\varepsilon'^{\ast} \leftrightarrow \varepsilon$
and $\nu \rightarrow -\nu$, and therefore $f$ is even and $g$ odd as function of $\nu$.
The amplitudes $f$ and $g$ can be determined by scattering circularly polarized photons
off nucleons polarized along or opposite to the photon momentum.
If the spins are parallel, the excited state must have spin $J\geq 3/2$, and therefore the transition can take place
only on a correlated three-quark system. The case of opposite spins is helicity conserving and
can take place on a single quark. Denoting the respective Compton tensors by
$T_{3/2}$ and $T_{1/2}$, we have $f(\nu)=(T_{1/2}+T_{3/2})/2$ and $g(\nu)=(T_{1/2}-T_{3/2})/2$. Analogous definitions
for the helicity-dependent cross sections lead to the total absorption cross section
$\sigma_T=(\sigma_{1/2}+\sigma_{3/2})/2$ and the transverse-transverse cross section
$\sigma_{TT}=(\sigma_{1/2}-\sigma_{3/2})/2$.
\newline \indent
The unitarity of the scattering matrix relates the absorption cross sections to the imaginary parts of the respective
forward scattering amplitudes by the optical theorem,
\begin{equation}
\label{eq1.2}
\rm{Im}\, f(\nu) = \frac{\nu}{4\pi}\,\sigma_T (\nu)\, , \quad
\rm{Im}\, g(\nu) = \frac{\nu}{4\pi}\,\sigma_{TT} (\nu)\, .
\end{equation}
By use of the
crossing relation and the optical theorem, the amplitudes can be expressed by
dispersion integrals,
\begin{equation}
\label{eq1.3}
{\rm{Re}}\, f(\nu) = f(0)+\frac{\nu^2}{2\pi^2}{\mathcal{P}}
\int_{\nu_0}^{\infty}\frac{\sigma_T(\nu')}
{\nu'^2-\nu^2}d\nu', \quad
{\rm{Re}}\, g(\nu) = \frac{\nu}{2\pi^2}{\mathcal{P}}
\int_{\nu_0}^{\infty}\frac{\nu'\sigma_{TT}(\nu')}
{\nu'^2-\nu^2} d\nu',
\end{equation}
where $\nu_0=m_{\pi}+m_{\pi}^2/2M\approx150$~MeV is the threshold of pion
photoproduction. Because $\sigma_T(\nu)$ is essentially constant for large $\nu$,
the dispersion relation (DR) for $f(\nu)$ has been subtracted at $\nu=0$ and the Thomson amplitude $f(0)$
appears as subtraction constant. On the other hand, recent data suggest that $g(\nu)$ obeys an unsubtracted DR.
\newline \indent
For $\nu<\nu_0$, the amplitudes of Eq.~(\ref{eq1.3}) are real and can be expanded
as a Taylor series in $\nu$. This series can be compared to the low-energy theorem (LET) of
Low~\cite{Low:1954kd} and Gell-Mann and Goldberger~\cite{GellMann:1954kc}. The LET fixes
the leading and next-to-leading terms by global
properties of the nucleon, that is, its mass $M$, charge $e_N$ ($e_p=1, e_n=0$), and anomalous magnetic moment
$\kappa_N$ ($\kappa_p=1.79, \kappa_n=-1.91$),
\begin{equation}
f(\nu) = -\frac{e^2e_N^2}{4\pi M} + (\alpha+\beta)
\nu^2+ {\mathcal{O}}(\nu^4) , \quad
g(\nu) = -\frac{e^2\kappa^2_N}{8\pi M^2}\,\nu +
\gamma_0\nu^3 + {\mathcal{O}}(\nu^5).
\label{eq1.5}
\end{equation}
As a result the internal structure (spectrum and excitation strength) of the complex
system becomes visible only through terms of relative order $\nu^2$. These terms
contain information on the dipole polarizabilities of the system, the forward scalar polarizability $\alpha+\beta$
and the forward spin polarizability $\gamma_0$. The next order in the expansion
is of relative order $\nu^4$ and contains contributions from
dipole retardation and higher multipoles.
\newline \indent
The comparison of Eq.~(\ref{eq1.5}) with the Taylor expansion of Eq.~(\ref{eq1.3}) yields
\begin{eqnarray}
\alpha + \beta & = & \frac{1}{2\pi^2}\,
\int_{\nu_0}^{\infty}\,\frac{\sigma_T(\nu')}{\nu'^2}\,d\nu' \, ,
\label{eq1.7} \\
\frac{\pi e^2\kappa^2_N}{2M^2} & = & \int_{\nu_0}^{\infty}\,\frac{\sigma_{3/2}(\nu')
-\sigma_{1/2}(\nu')}{\nu'}\,d\nu' \, \equiv I_{\rm {GDH}} \, , \label{eq:1.8}\\
\gamma_0 & = & -\frac{1}{4\pi^2}\,\int_{\nu_0}^{\infty}\,
\frac{\sigma_{3/2}(\nu')-\sigma_{1/2}(\nu')}
{\nu'^3}\,d\nu \, ,\label{eq1.9}
\end{eqnarray}
which are the sum rules of Baldin~\cite{Baldin:1960dd},
Gerasimov, Drell, and Hearn~\cite{Gerasimov:1965et,Drell:1966jv}, and Gell-Mann,
Goldberger, and Thirring~\cite{GellMann:1954kc,GellMann:1954db}, in order.

\begin{figure}
\includegraphics[width=.5\textwidth, angle=90]{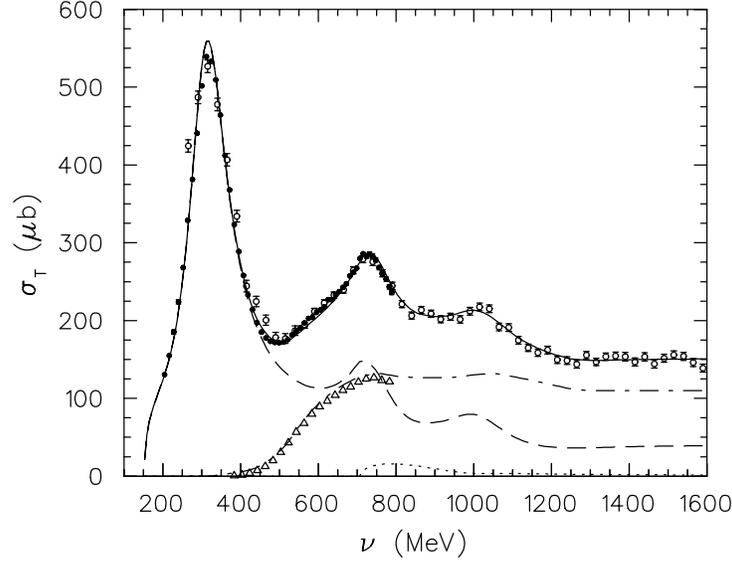}
\caption{The total absorption cross section $\sigma_T$ for the proton as function
of the photon lab energy $\nu$, in units of $\mu$b. Data from
MAMI~\cite{MacCormick:1996jz} (solid circles) and Daresbury~\cite{Armstrong:1971ns} (open circles), the
triangles represent the $2\pi$ contributions measured at MAMI. The lines are
MAID results~\cite{Drechsel:1998hk} for the total cross section (solid),
one-pion channels (dashed), more-pion channels (dash-dotted), and
$\eta$ channel (dotted). Figure from Ref.~\cite{Drechsel:2004ki}.}
\label{fig_sig}
\end{figure}
%
\begin{figure}
\includegraphics[width=.5\textwidth, angle=90]{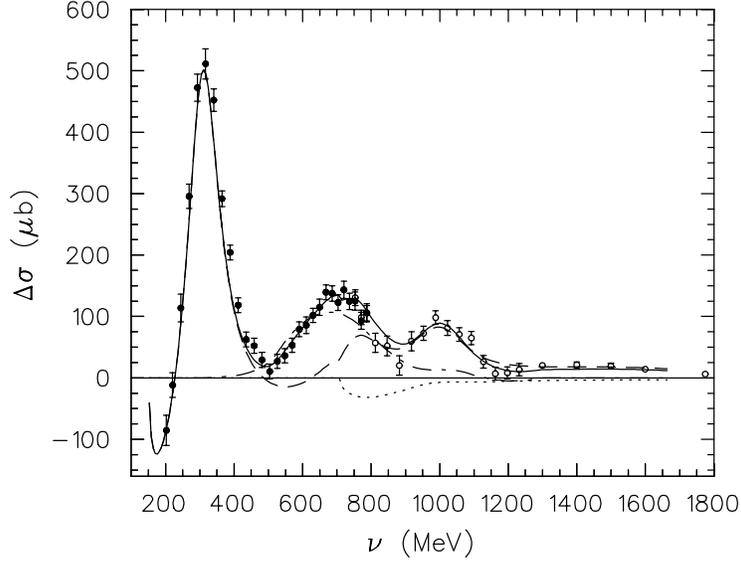}
\caption{The helicity
difference $\Delta\sigma = \sigma_{3/2}-\sigma_{1/2}$ for the proton as function
of the photon lab energy $\nu$, in units of $\mu$b.
Data from MAMI~\cite{Ahrens:2001qt} (solid circles) and ELSA~\cite{Dutz:2003mm} (open circles).
The lines represent MAID results, see Fig.~\ref{fig_sig} for notation. Figure from Ref.~\cite{Drechsel:2004ki}.}
\label{fig_Dsig}
\end{figure}
The total absorption cross section $\sigma_T$ is displayed in Fig.~\ref{fig_sig}, together
with the contributions of individual reaction channels. The cross section is characterized
by a shoulder near threshold (non-resonant pion production), the large peak of the $P_{33}(1230)$
resonance, and smaller peaks in the second and third resonance regions. The two-pion
channels provide a large background for energies $\nu>700$~MeV.
Above $\nu\approx2$~GeV, the cross section decreases to about $130~\mu$b, and at the highest measured energies
a weak logarithmic increase was observed. Obviously, an unsubtracted dispersion
relation does not exist. The situation is quite different for $\Delta \sigma=\sigma_{3/2}-\sigma_{1/2}$
plotted in Fig.~\ref{fig_Dsig}, which shows a much reduced background and only small helicity differences
for $\nu>1.3$~GeV. We may therefore assume that (i) the high-energy photon is ``helicity blind''
and (ii) an unsubtracted DR exists.
\newline \indent
The one-pion contribution to $\Delta \sigma$ has the multipole decomposition
\begin{eqnarray}
\label{eq1.10}
[\sigma_{3/2}-\sigma_{1/2}]_{1\pi}
& = &  8\pi \frac{k_{\pi}} {k_{\gamma}}\, \bigg( -|E_{0+}|^2 + |M_{1+}|^2 - 6 {\rm{Re}}\,
(E_{1+}^{\ast}M_{1+}) - 3|E_{1+}|^2  \nonumber\\
&&- |M_{1^-}|^2 + |E_{2-}|^2 + 6\ \mbox{Re}\ (E_{2-}^{\ast}M_{2-}) - 3|M_{2^-}|^2 \pm \ ...
\bigg) \, ,
\end{eqnarray}
with $k_{\gamma}$ and $k_{\pi}$ the momenta of photon and pion, respectively.
Near threshold the pions are produced in the S wave (spin $J=1/2$, multipole $E_{0+}$), that is,
$\Delta \sigma<0$. In the region of
the $P_{33}(1232)$ with $J=3/2$, both helicity cross
sections contribute, but since $M_{1+}$ dominates, we have
$\sigma_{3/2}\approx 3\sigma_{1/2}$ and $\Delta \sigma$ is positive and large.
According to Fig.~\ref{fig_Dsig}, $\sigma_{3/2}$ dominates the photoabsorption cross section
also in the second and third resonance regions. It was one of
the early successes of the quark model to predict the small value of $\sigma_{1/2}$  by
cancelation of convection and spin currents. The GDH collaboration extended the measurement
up to 2.9~GeV at ELSA and found that $\sigma_{3/2}-\sigma_{1/2}$ turns
to negative values at $\nu\approx2$~GeV, as was also predicted by extrapolation of
DIS data~\cite{Bianchi:1999qs,Simula:2001iy}.
\begin{table}[h,t]
\label{tab_GDH}
\begin{tabular}{|l|cc|c|}
\hline
$\nu$ [GeV] &  $I_{\rm {GDH}}^p[\mu$b]  & $\gamma_0^p\, [10^{-4}~{\rm {fm}}^4]$ &Ref.  \\
\hline
$<0.2$   &  $-27.5\pm3$        &  $0.90\pm0.05$  &\cite{Drechsel:1998hk,Arndt:2002xv} \\
$0.2-0.8$    &  $226\pm5\pm12$     &  $-1.87\pm0.08\pm0.10$  &\cite{ Ahrens:2001qt}\\
$0.8-2.9$     &  $27.5\pm2.0\pm1.2$ & $-0.03$   & \cite{Dutz:2004zz} \\
\hline
total       &  $226 \pm6 \pm12$    &  $-1.00 \pm0.08 \pm 0.11$   & \\
\hline
sum rule    & $204$                  &  --    &   \\
\hline
\end{tabular}
\caption{The GDH sum rule $I_{\rm {GDH}}^p$ and the forward spin polarizability $\gamma_0^p$
for the proton.}
\label{tab:GDH}
\end{table}
\subsection{GDH sum rule and forward spin polarizability}
The helicity dependent cross sections were measured by the GDH Collaboration at MAMI and ELSA,
in the energy range $0.2-2.9$~GeV for the proton and $0.2-1.9$~GeV for the neutron ($^2$H target).
Based on a LET for pion photoproduction and several data, the MAID and SAID analyses provide reliable estimates
for the threshold region below $0.2$~GeV. Table~\ref{tab:GDH} shows that the resulting
GDH integral up to $2.9$~GeV exceeds the
sum rule value by about 10\%. However, Regge estimates for the tail above $2.9$~GeV yield an
additional contribution of $(-14 \pm 2)\mu$b~\cite{Bianchi:1999qs,Simula:2001iy}. Furthermore, the
LEGS Collaboration has recently remeasured the $(\gamma,\pi^0)$ reaction in the $\Delta(1232)$ region
and extracted a smaller contribution than obtained at MAMI, which would further reduce the integral by
$(-18 \pm 6)\mu$b~\cite{Hoblit:2008iy}. In conclusion, the GDH integral for the proton
is essentially saturated by the threshold and resonance regions with only a small contribution of order
5\% from the Regge tail.
\newline \indent
A simple analysis of the deuteron data~\cite{Dutz:2005ns,Ahrens:2006yx} also confirms
the validity of the GDH sum rule for the neutron, albeit with much larger error bars. It remains a
fundamental problem how to extract the neutron properties from a nuclear target, for example, how to ``divide''
coherent $\pi^0$ production and proton-neutron break-up between the nuclear and nucleonic degrees
of freedom. Future experiments to determine the sum rule also for the ``neutron target'' $^3{\rm {He}}$ will
provide a better understanding of this problem.
\newline \indent
Table~\ref{tab:GDH} also lists the experimental results for the forward spin polarizability $\gamma_0$
of the proton. Because of the additional factor $1/\nu^2$ relative to $I_{\rm {GDH}}$, the threshold region
becomes very important, whereas the high-energy contribution is negligible. The resulting delicate cancelation
of threshold and resonance contributions poses
a big problem for ChPT and phenomenological models (see Table~10 of Ref.~\cite{Drechsel:2002ar}). On the other
hand, the precision experiment of the GDH Collaboration was a first step to disentangle the 4
spin (dipole) polarizabilities of the proton. At present, only the backward combination is known
from RCS, although with a much larger error bar. Further progress requires polarization degrees of
freedom~\cite{Pasquini:2007hf}, experimental projects along these lines are under consideration.
\section{Forward VVCS and Electroproduction}
\subsection{VVCS formalism}
\begin{figure}
\includegraphics[width=.7\textwidth, angle=90]{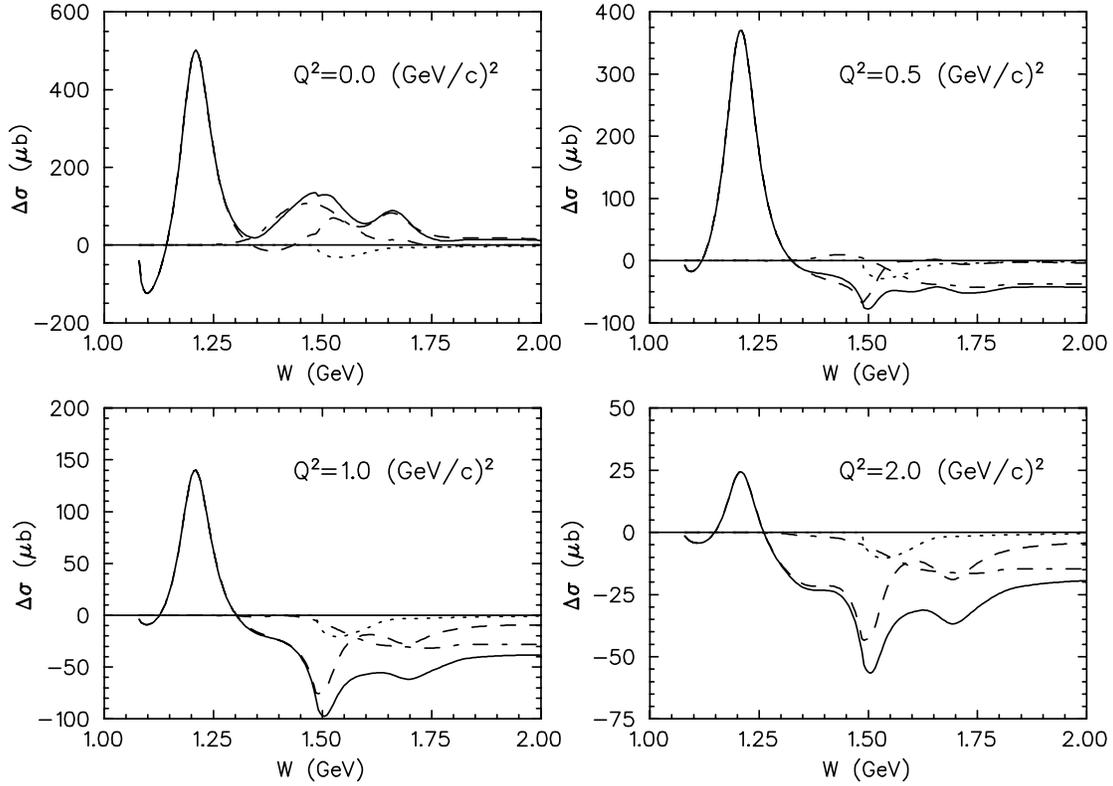}
\caption{The helicity difference $\Delta\sigma =\sigma_{3/2}-\sigma_{1/2}=-2\sigma_{TT}$ for
the proton as a function of c.m. energy $W$ and for momentum transfers
$Q^2=0, 0.5 ,1.0$, and $2.0~{\rm {GeV}}^2$ according to MAID~\cite{Drechsel:1998hk}. The lines
shows the total result (solid) and the contributions of
one-pion (dashed), more-pion (dash-dotted), and $\eta$
(dotted) production. Figure from Ref.~\cite{Drechsel:2004ki}.}
\label{fig_DsigQ2}
\end{figure}

\begin{figure}
\includegraphics[width=.7\textwidth]{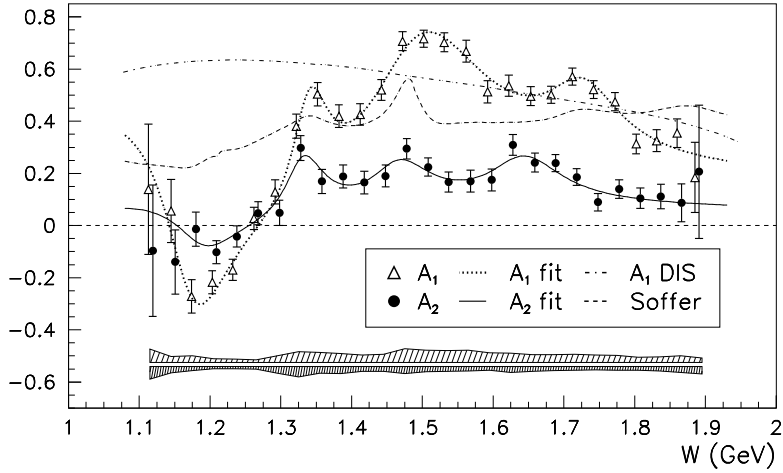}
\caption{The photon asymmetries for the proton as a function of the c.m. energy $W$ at momentum transfer
$Q^2\approx 1.33\,{\rm {GeV}}^2$. Data from the RSS Collaboration~\cite{Wesselmann:2006mw},
triangles: $A_1$, solid circles: $A_2$. Dotted and dashed lines: fits to $A_1$ and $A_2$,
respectively, dot-dashed line: DIS extrapolation to
$A_1$~\cite{Anthony:2000fn}, dashed line: phenomenological fit
to $A_1$~\cite{Soffer:2000zd}. Figure from Ref.~\cite{Wesselmann:2006mw}.}
\label{fig_A12p}
\end{figure}

In this section we consider the forward scattering of a virtual
photon $\gamma^{\ast}$ with momentum 4-vector $q$ on a nucleon $N$
with momentum 4-vector $p$. The Lorentz products of these 4-vectors
are $p \cdot p =M^2$, $q \cdot q =-Q^2$, and $p \cdot q =M \nu$,
with $Q^2$ the ``virtuality'' of the photon. The Lorentz scalar $\nu$ takes the
value of the photon lab energy, $\nu = E_{\gamma}^{\rm {lab}}$,
the invariant mass $W$ of the excited state is $W^2= 2M\nu+M^2-Q^2$.
The VVCS tensor $T=T(\nu,Q^2)$ has the form of a $2 \times 2$ matrix in nucleon spinor space,
\begin{equation}
\label{eq2.1}
T= {\vec{\varepsilon}}'^{\ast}\cdot{\vec{\varepsilon}}\; f_T + f_L
+i{\vec{\sigma}}\cdot({\vec{\varepsilon}}'^{\ast}\times {\vec{\varepsilon}})\; g_{TT}
- i{\vec{\sigma}}\cdot[({\vec{\varepsilon}}'^{\ast}-{\vec{\varepsilon}})\times \hat{q}]\;  g_{LT}\, ,
\end{equation}
with ${\vec{\varepsilon}}\,'$ and ${\vec{\varepsilon}}$ the transverse photon
polarizations and $\hat{q}$ the longitudinal one. Because of the crossing symmetry,
$g_{TT}=g_{TT}(\nu,Q^2)$ is an odd function of $\nu$, the other 3 amplitudes are even functions of $\nu$.
The optical theorem relates the imaginary parts of the 4 amplitudes to the
4 partial cross sections of inclusive scattering,
\begin{equation}
\label{eq2.2}
{\mbox{Im}}\, \{ f_T, f_L, g_{TT}, g_{LT}\} =
\frac{K}{4\pi}\, \{\sigma_T, \sigma_L, \sigma_{TT}, \sigma_{LT}\}\,,
\end{equation}
with $K=K(\nu,Q^2)$ the ``equivalent photon energy''.
We note that products such as $K\,\sigma_T$ are independent of
the choice of $K$, because they are directly proportional to
the measured cross section.
For further use we also list the relations between the inclusive cross sections and the
nucleon structure functions used to describe DIS,
\begin{equation}
\label{eq2.2DIS}
\{\sigma_T, \sigma_L, \sigma_{TT}, \sigma_{LT}\}= C \left \{ F_1,\;
\frac {M(1+\gamma^2)}{\nu \gamma^2}F_2-F_1,\; g_1-\gamma^2 g_2,\; \gamma(g_1+g_2) \right \}\,,
\end{equation}
with $\gamma=Q/\nu$ and a common factor $C=4 \pi^2 \alpha_{em}^2/MK$.
The spin-independent amplitudes $f_T$ and $f_L$ (or the structure functions $F_1$ and $F_2$)
are of interest in their own right, however, we concentrate on the
spin-dependent amplitudes in the following.
The imaginary parts of the scattering amplitudes get contributions from the Born terms
(poles at $\nu= \pm \nu_B(Q^2)=\pm Q^2/2M$) and from inelastic processes above pion threshold
($\nu > \nu_0(Q^2)=m_{\pi}+m_{\pi}/2M+\nu_B(Q^2)$). The Born contributions are
\begin{equation}
\label{eq2.3}
g_{TT}^B = -\frac{\alpha_{em}\nu}{2M^2}
\left (F_P^2 + \frac{Q^2 G_M^2}{\nu^2-\nu_B^2+i \epsilon}\right), \quad
g_{LT}^B = \frac{\alpha_{em}Q}{2M^2}
\left (F_DF_P - \frac{Q^2 G_EG_M }{\nu^2-\nu_B^2+i \epsilon}\right),
\end{equation}
with $G_E$ the electric and $G_M$ the magnetic
Sachs form factors, $F_D$ the Dirac form factor, and
$F_P$ the Pauli form factor of the nucleon. Note that the limits
$Q^2 \rightarrow 0$ and $\nu \rightarrow 0$ can not be interchanged
in Eq.~(\ref{eq2.3}). In particular, the real photon limit is obtained by first
choosing $Q^2\rightarrow0$. As a result the Born contribution to real photon scattering
is real, because the real photon can not be absorbed on an nucleon. To the
contrary, the virtual photon is absorbed in elastic electron scattering, as
is expressed by the pole terms appearing in Eq.~(\ref{eq2.3}).
These pole terms fulfill DRs by themselves.
The remaining non-pole or dispersive amplitudes fulfill the DRs
\begin{eqnarray}
\label{eq2.4}
{\rm{Re}} g_{TT}^{\rm{disp}}(\nu,\,Q^2) &=&
\frac{\nu}{2\pi^2}\,{\mathcal{P}}\,
\int_{\nu_0}^{\infty}\frac{K(\nu',Q^2)\;\sigma_{TT}(\nu',Q^2)}
{\nu'^2-\nu^2} d\nu' \, ,\\
{\rm{Re}} g_{LT}^{\rm{disp}}(\nu,Q^2)&=&
\frac{1}{2\pi^2}\,{\mathcal{P}}\,
\int_{\nu_0}^{\infty}\frac{K(\nu',Q^2)\; \nu' \sigma_{LT}(\nu',Q^2)}
{\nu'^2-\nu^2} d\nu' \,.
\label{eq2.4LT}
\end{eqnarray}
For $\nu<\nu_0$, these dispersive amplitudes are real and
given by power series,
\begin{eqnarray}
\label{eq2.5}
{\rm{Re}} g_{TT}^{\rm{disp}}(\nu,\,Q^2) &=&
\frac{2 \, \alpha_{em} }{M^2}  \, I_{TT}(Q^2) \nu
+\gamma_{TT}(Q^2) \nu^3 + {\mathcal{O}}(\nu^5) \, ,\\
{\rm{Re}} g_{LT}^{\rm{disp}}(\nu,\,Q^2) &=&
\frac{2 \, \alpha_{em}}{M^2}  \, Q  I_{LT}(Q^2)
+ Q \delta_{LT}(Q^2) \nu^2 + {\mathcal{O}}(\nu^4) \, .
\label{eq2.5LT}
\end{eqnarray}
Equations~(\ref{eq2.4})- (\ref{eq2.5LT}) yield the following sum rules:
\begin{eqnarray}
I_{TT}(Q^2) &=&
\frac{M^2}{\pi \, e^2}\,
\int_{\nu_0}^{\infty} \frac{K(\nu, Q^2)\,\sigma_{TT}(\nu,Q^2)}{\nu^2}
d\nu \, ,\label{eq2.7} \\
\gamma_{TT}(Q^2) &=& \frac{1}{2\pi^2}\,
\int_{\nu_0}^{\infty} \frac{K(\nu, Q^2)\,\sigma_{TT}(\nu,Q^2)}{\nu^4}
d\nu \, ,\label{eq2.8}\\
I_{LT}(Q^2) &=&
\frac{M^2}{\pi \, e^2}\,
\int_{\nu_0}^{\infty} \frac{K(\nu, Q^2)\; \sigma_{LT}(\nu,Q^2)}{\nu Q}
d\nu \, , \label{eq2.7LT}\\
\delta_{LT}(Q^2) &=& \frac{1}{2\pi^2}\,
\int_{\nu_0}^{\infty} \frac{K(\nu, Q^2)\; \sigma_{LT}(\nu,Q^2)}{\nu^3 Q}
d\nu \, .
\label{eq2.8LT}
\end{eqnarray}
with $I_{TT}(0)=-(M^2/2\pi e^2)\, I_{\rm {GDH}}$ and $\gamma_{TT}(0)=\gamma_0$.
For $Q^2 \rightarrow 0$, $\sigma_{LT}/Q$ is finite and also $I_{LT}(0)$ and $\delta_{LT}(0)$
exist, although they can not be determined with real photons.
\newline \indent
The one-pion contribution to $\Delta\sigma$ of the proton is displayed in
Fig.~\ref{fig_DsigQ2} as function of the c.m. energy $W$ for several
momentum transfers. The figure shows that both the pion S-wave production near threshold
and the $P_{33}(1232)$ resonance contribution decrease rapidly with $Q^2$.
A striking feature is seen in the second and third resonance regions,
for which $\Delta\sigma$ changes sign at $Q^2\approx  0.3 \,{\rm{GeV}}^2$. For larger $Q^2$
the high energy contributions increase in strength relative to
the first resonance region. Equation~\ref{eq1.10} explains one of the reason why this happens.
The photoexcitation of the $D_{13}(1520)$ is dominated by electric dipole radiation
(multipole $E_{2-}$). However, magnetic quadrupole radiation
(multipole $M_{2-}$) becomes more and more important with increasing virtuality of the photon.
The results are the observed sign change of $\Delta\sigma$ and
$\sigma_{1/2}$ dominance if $Q^2$ increases further. Finally, if $Q^2$ reaches values beyond $4 \,{\rm {GeV}}^2$,
the resonance structures become small fluctuations on top of the low-energy tail of DIS.
\newline \indent
%
Information on the longitudinal strength and the amplitude $\sigma_{LT}$ is still scarce. Considerable progress
has been made by the RSS Collaboration who measured both spin-dependent cross sections near $Q^2=1.3\,{\rm {GeV}}^2$.
Figure~\ref{fig_A12p} shows the measured photon asymmetries as function of the c.m. energy $W$.
These asymmetries are related to the cross sections by $A_1=\sigma_{TT}/\sigma_{T}$ and $A_2=\sigma_{LT}/\sigma_{T}$.
Several resonance structures are clearly visible in both asymmetries. We conclude that the total absorption
cross section $\sigma_{T}$ contains a much stronger non-resonant background than is the case for the
spin-dependent cross sections $\sigma_{TT}$ and $\sigma_{LT}$. As a result we may argue that integrals over the latter
cross sections are essentially saturated by the resonance region below $W=2$~GeV.
The structure of $\sigma_{TT}=-\textstyle {\frac {1}{2}}\Delta\sigma$ is as found in the previous
Fig.~\ref{fig_DsigQ2} for the one-pion channel, except for the region $W=(1.35-1.5)$~GeV in which large contributions from two-pion channels are
expected. The amplitude $\sigma_{LT}$ has a similar peak structure, however, in comparison with $\sigma_{TT}$ it appears suppressed.
Moreover, the higher peaks are shifted to lower energies. The sharp peak of both amplitudes near $W=1.34$~GeV is
somewhat surprising. The closest resonance, $P_{11}(1440)$, has a large width and also the two-pion contribution is
expected to yield a broad structure, see Figs.~\ref{fig_sig} and \ref{fig_Dsig}.
\newline \indent
%
The one-pion contribution to $\sigma_{LT}$ has the following multipole expansion:
\begin{eqnarray}
\label{eq2.9}
[\sigma_{LT}]_{1\pi}
& = &  4\pi \frac{k_{\pi}Q} {k_{\gamma}K_W}\, {\rm {Re}} [S_{0+}^{\ast}E_{0+} + 2S_{1+}^{\ast}(3E_{1+}+M_{1+}) \nonumber\\
 && - S_{1-}^{\ast}M_{1-}+2S_{2-}^{\ast}(E_{2-}-3M_{2-})\pm \cdots]\ ,
\end{eqnarray}
with $K_W=(M/W)K$. A comparison of Fig.~\ref{fig_A12p} with Eq.~(\ref{eq2.9}) yields the relative phases of the
charge and transverse multipoles, more precisely the relative phase of the helicity amplitudes, $S_{1/2}^{\ast}A_{1/2}$.
The immediate result is $S_{0+}^{\ast}E_{0+}>0$ near threshold, $S_{1+}^{\ast}M_{1+}<0$ near the $P_{33}(1232)$, and
$S_{2-}^{\ast}M_{2-}<0$ near the $D_{13}(1520)$. The peak near $W=1.63$~GeV may be due to the  $S_{31}(1620)$,
which has large values and same signs for $S_{0+}$ and $E_{0+}$.
\subsection{VVCS sum rules, polarizabilities, and higher twists}
Whereas Eqs.~(\ref{eq2.7}) and (\ref{eq2.7LT}) are appropriate definitions in the resonance region,
asymptotic QCD is described by integrals $I_{ \{1,2\} }$ over the nucleon structure functions,
\begin{equation}
\label{eq4.1}
I_{ \{1,2\} }(Q^2)=\frac {2M^2}{Q^2} \int_{0}^{x_0}\, g_{ \{1,2\}}(x,Q^2)\, dx \equiv \frac {2M^2}{Q^2}\Gamma_{ \{1,2\}}(Q^2) \,,
\end{equation}
with $\Gamma_{ \{1,2 \} }$ the inelastic contribution to the first moment of the respective structure function and
$x=Q^2/2M\nu$ the Bjorken variable. Assuming a ``superconvergence relation'', Burkhardt and
Cottingham~\cite{Burkhardt:1970ti} postulated that the inelastic
contribution to the first moment of $g_2$ is canceled by the elastic contribution, which leads to the relation
\begin{equation}
\label{eq4.2}
I_2(Q^2) \equiv \frac{2M^2}{Q^2}\int_0^{x_0}g_2(x,Q^2)\,dx
=\frac{1}{4} \, F_P(Q^2) \, \left( F_D(Q^2) + F_P(Q^2) \right) \, .
\end{equation}
This result is astounding, because it relates the excitation spectrum to ground state properties
at all distances. If the BC sum rule holds, the discussed integrals take the following values at the real photon point:
\begin{equation}
\label{eq4.2RCS}
I_1(0)=I_{TT}(0)= -\frac{1}{4}\kappa_N^2\, ,\quad I_2(0) = \frac{1}{4}\kappa_N (e_N + \kappa_N) \, ,
\quad I_{LT} (0) =\frac{1}{4} e_N \kappa_N.
\end{equation}
%
\begin{figure}
\includegraphics[width=.45\textwidth]{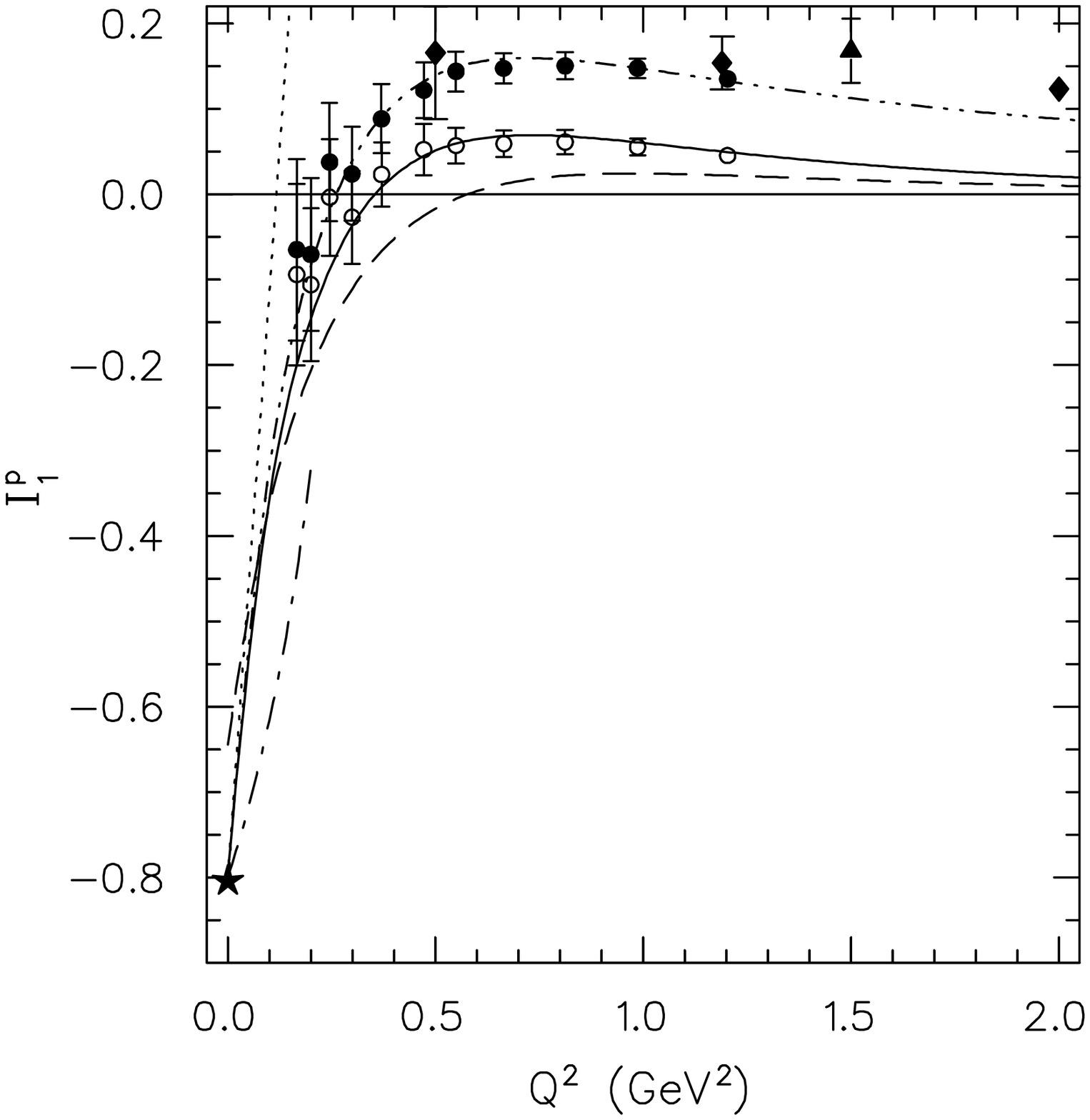}
\includegraphics[width=.55\textwidth]{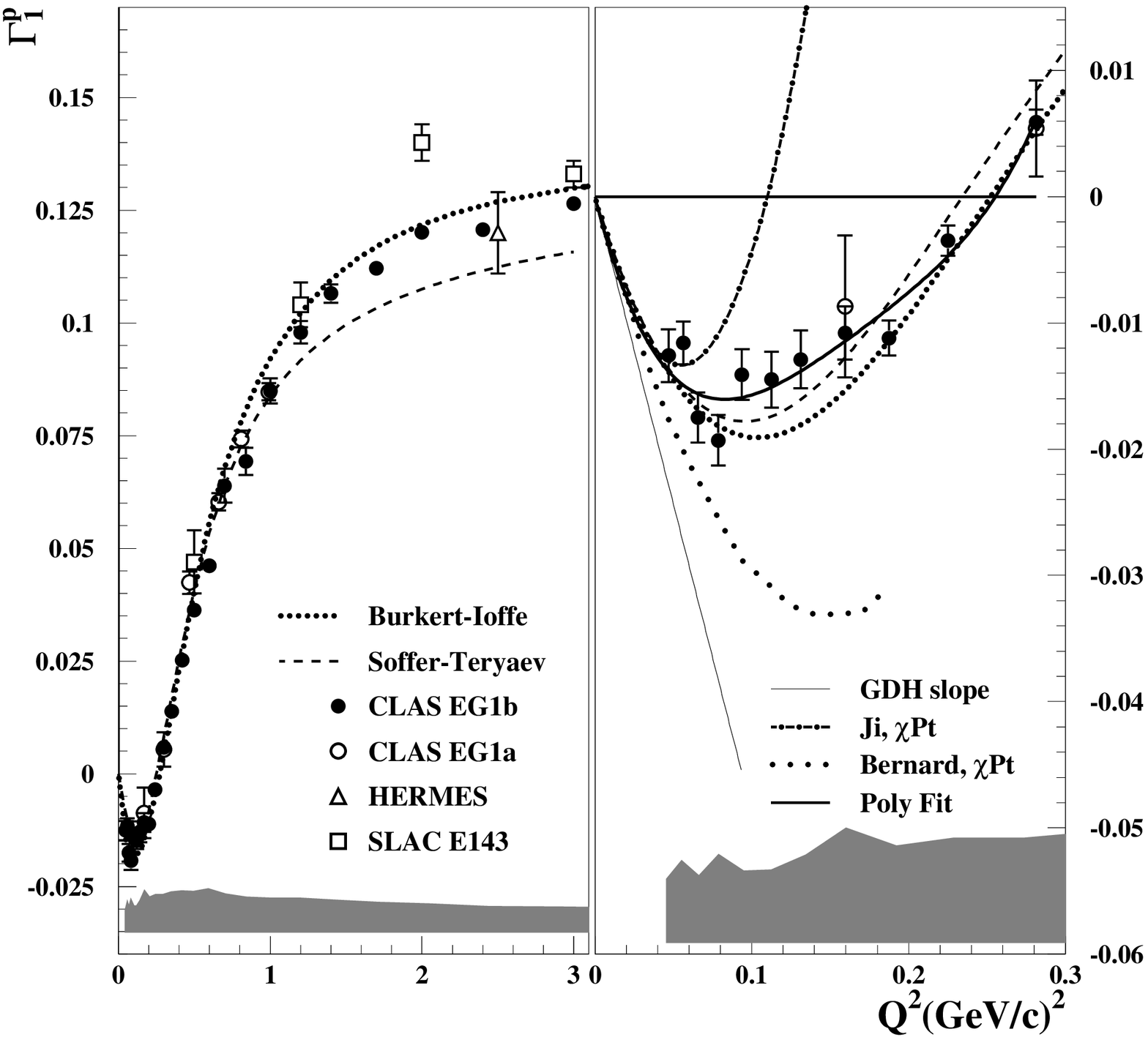}
\caption{Left: $I_1^p$ as function of $Q^2$. Data from the CLAS Collaboration~\cite{Fatemi:2003yh},
measured resonance contribution for $W<2$~GeV (open circles) and with DIS contribution (solid circles).
Further data from SLAC~\cite{Abe:1998wq} (diamonds) and HERMES~\cite{Airapetian:2002wd} (triangles).
Dashed line: one-pion channel integrated up
to $W=2$~GeV, full line: with two-pion and eta channels included~\cite{Drechsel:2002ar}. Dotted lines:
${\mathcal{O}}(p^4)$ predictions of heavy baryon
ChPT~\cite{Ji:1999sv,Kao:2002cp}, dash-dotted lines relativistic baryon ChPT
~\cite{Bernard:2002bs,Bernard:2002pw}, dash-dot-dotted line: interpolating formula, asterisk: GDH sum rule.
Figure from Ref.~\cite{Drechsel:2004ki}.
\newline
Right: $\Gamma_1^p$ as function of $Q^2$. Data from the CLAS Collaboration~\cite{Prok:2008ev} with DIS
contributions included (solid circles) and earlier CLAS~\cite{Fatemi:2003yh}, HERMES~\cite{Airapetian:2007mh},
and SLAC~\cite{Abe:1996ag} data,
compared to phenomenological models (dashed,~\cite{Burkert:1992tg} and dotted,~\cite{Soffer:2004ip}),
HBChPT (dot-dashed,~\cite{Ji:1999pd}), and  relativistic ChPT~(lower dotted line in right
inset,~\cite{Bernard:2002pw}). Figure from Ref.~\cite{Prok:2008ev}.}
\label{fig_I1p}
\end{figure}
The integral $I_1^p$ and the related moment $\Gamma_1^p$ are displayed in Fig.~\ref{fig_I1p}. In the left inset,
the CLAS data are compared with MAID~\cite{Drechsel:2002ar}. The full line including one- and two-pion channels
as well as the eta describes the measured contribution up to $W=2$~GeV~(open circles, \cite{Fatemi:2003yh}) reasonably well,
the one-pion contribution (dashed line) fails already at relatively small momentum transfer. Also the DIS
contributions become increasingly important at larger values of $Q^2$, see the dash-dot-dotted line and the solid circles.
The right panel shows a comparison of the data with ChPT predictions. It is evident that the predictions fail
for $Q^2>0.05~{\rm {GeV}}^2$, probably because of vector-meson and resonance contributions.
The agreement is expected to improve for the isovector combination $\Gamma_1^p-\Gamma_1^n$,
because the large $P_{33}(1232)$ contributions cancel in this expression. The experimental data have been fitted to the form
\cite{Deur:2008ej}
\begin{equation}
\label{eq4.3}
\Gamma_1^{p-n}(Q^2)= \frac {(\kappa_n^2-\kappa_p^2)Q^2}{8M^2}+a \frac {Q^4}{M^4}+b \frac {Q^6}{M^6}+\ldots \,,
\end{equation}
with the results $a=0.62 \pm 0.05 \pm 0.18$ and $b=0.77 \pm 0.11 \pm 0.27$. The prediction of HBChPT at ${\mathcal {O}}(p^4)$
agrees, $a=0.58$~\cite{Ji:1999pd}, whereas Lorentz invariant BChPT ${\mathcal {O}}(p^4)$ yields a much larger value
$a=1.87$~\cite{Bernard:2002pw}.
\newline \indent
%
\begin{figure}
\includegraphics[width=.55\textwidth]{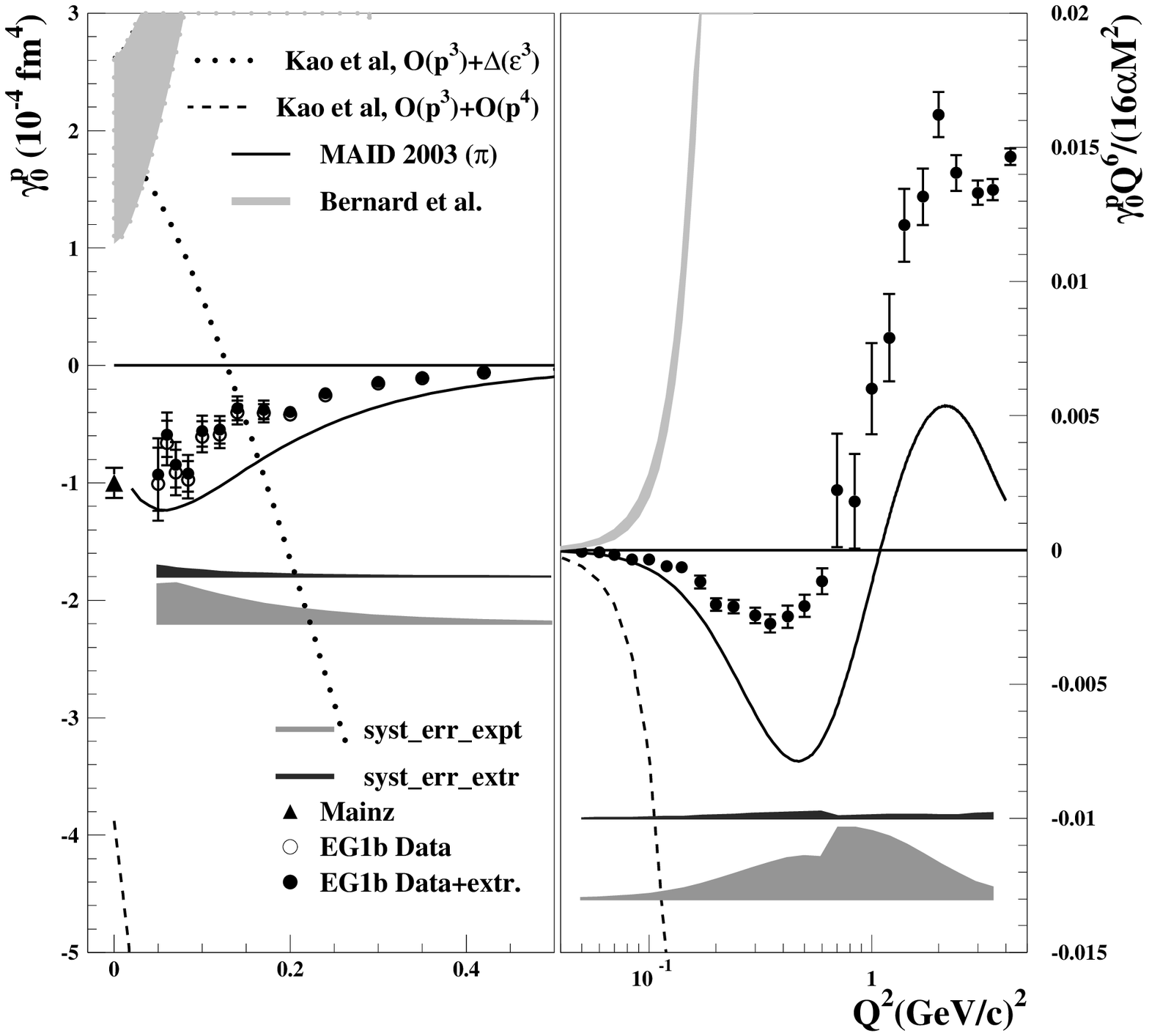}
\includegraphics[width=.45\textwidth]{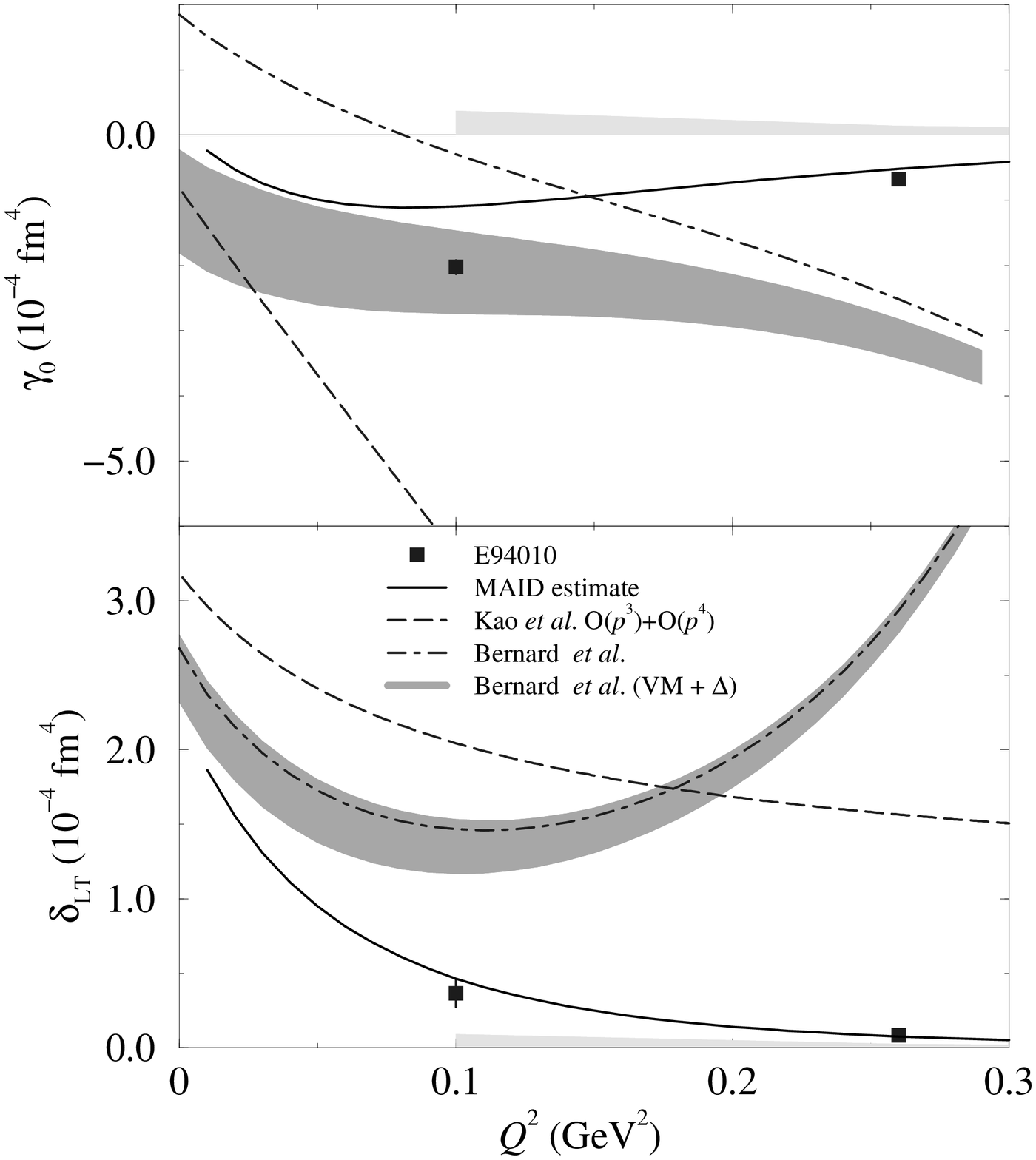}
\caption{The generalized spin polarizabilities as function of $Q^2$.
Left panels: $\gamma_{TT}^p$, value at $Q^2=0$ (triangle) measured at MAMI and  ELSA~\cite{Dutz:2003mm},
CLAS data~\cite{Prok:2008ev} at finite $Q^2$ (open circles: measured, full circles:
including extrapolation to DIS). Solid line: the one-pion contribution of MAID~\cite{Drechsel:1998hk},
dotted and dashed lines: HBChPT at
${\mathcal {O}}(p^4)$ and ${\mathcal {O}}(\epsilon^3)~$\cite{Kao:2002cp},
grey error band (top left): relativistic ChPT~\cite{Bernard:2002pw}. Figure from Ref.~\cite{Prok:2008ev}.
\newline
Right panel: $\gamma_{TT}^n$ (top) and $\delta_{LT}^n$ (bottom), data from E94010~\cite{Amarian:2004yf}.
Solid lines: MAID~\cite{Drechsel:1998hk}, dashed lines: HBChPT at
${\mathcal {O}}(p^4)$~\cite{Kao:2002cp}, error bands and dash-dotted lines: relativistic ChPT~\cite{Bernard:2002pw}
with and without $\Delta$ and vector meson contributions. Figure from Ref.~\cite{Amarian:2004yf}.}
\label{fig_gamma}
\end{figure}
Recent experimental results for the forward spin polarizabilities (FSPs) of proton and neutron are shown in Fig.~\ref{fig_gamma}.
The left panel compares $\gamma_{TT}^p(Q^2)$ with ChPT and MAID in the low-$Q^2$ region. Whereas MAID
reproduces the values close to the photon point, it misses the data at larger momentum transfers. As in the case
of real photons, the predictions of ChPT miss the data completely. In the central panel, $\gamma_{TT}^p$ is multiplied
with a factor $Q^6$, because this product approaches the third moment of the structure function $g_1$ at large values
of $Q^2$. In the limit of Bjorken scaling, this moment should approach a plateau, which may have been actually reached
at $Q^2=4-5~{\rm {GeV}}^2$. These results are remarkable for the following reasons. The FSP at the
photon point is $-1$ (here and in the following in units of $10^{-4}~{\rm {fm}}^{-4}$), a very small value compared to the
RCS backward spin polarizability of about $-38$. The data points at the highest values $Q^2$ translate into FSPs
of about $3 \cdot 10^{-4}$, four orders of magnitude smaller than the RCS value, that is,
the generalized polarizabilities disappear rapidly if $Q^2$ approaches the scaling region.
The right panel in Fig.~\ref{fig_gamma} shows the two FSPs of the neutron, $\gamma_{TT}^n(Q^2)$ and $\delta_{LT}^n(Q^2)$.
The agreement with MAID is quite acceptable, in particular for $\delta_{LT}^n$ (lower part of the panel).
Contrary to all expectations, ChPT fails to describe $\delta_{LT}^n$ even at $Q^2=0.1~{\rm {GeV}}^2$.
\newline \indent
\begin{figure}
\includegraphics[width=.5\textwidth]{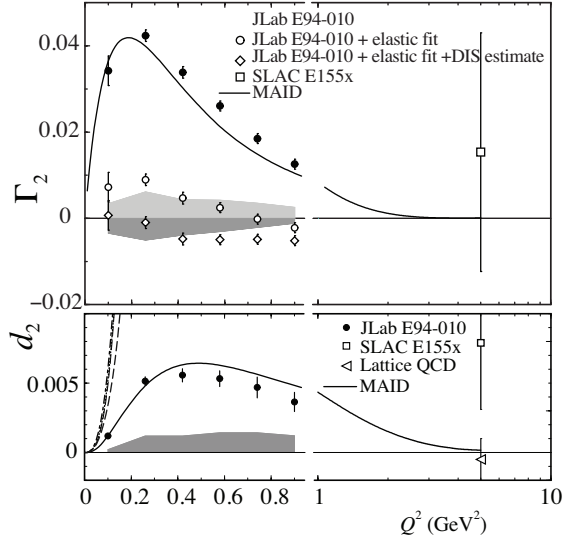}
\caption{Top: $\Gamma_2^n$ as function of $Q^2$. Inelastic contribution measured by the E94010 Collaboration
(solid circles,~\cite{Amarian:2003jy}) compared to MAID (solid line,~\cite{Drechsel:1998hk}),
open circles: measured data plus elastic contribution~\cite{Mergell:1995bf}, diamonds: measured data plus elastic plus DIS contributions.
\newline
Bottom: The inelastic contribution to the neutron moment $d_2^n$ as function of $Q^2$.
Data from JLab E94-010 (open circles,~\cite{Amarian:2003jy}) and SLAC E99-117/E155x (square,~\cite{Anthony:2002hy}),
compared to MAID (solid line,~\cite{Drechsel:1998hk}), HBChPT (dotted,~\cite{Kao:2002cp}),
relativistic ChPT (dash-dotted line,~\cite{Bernard:2002pw}), and a prediction of
lattice QCD~\cite{Gockeler:2000ja}
at $Q^2=5~{\rm {GeV}}^2$. Figure from Ref.~\cite{Amarian:2003jy}.
}
\label{fig_Gamma2n_d2n}
\end{figure}
The validity of the BC sum rule is demonstrated by the upper panel of Fig.~\ref{fig_Gamma2n_d2n} showing $\Gamma_2^n(Q^2)$.
The data of the E94010 Collaboration~\cite{Amarian:2003jy} (solid circles) are in reasonable agreement
with MAID~\cite{Drechsel:1998hk} (solid line), both integrated over the excitation spectrum from threshold
to $W=2$~GeV. Addition of the elastic contribution leads to the open circles, further addition of the DIS
estimate to the diamonds at slightly negative values. The observed cancelations support the existence of the
BC sum rule for the neutron. A recent result of the RSS Collaboration yields an even more striking result
for the proton. At $Q^2=1.28~{\rm {GeV}}^2$, the (positive) DIS contribution cancels 99\% of the
(negative) elastic and resonance contributions, the remainder
is far below the experimental error~\cite{Slifer:2008xu}.
\newline \indent
Up to this point we have concentrated on GDH-like integrals and polarizabilities, which are
defined by low energy expansions of VVCS amplitudes.
Starting from asymptotic QCD, the operator product expansion describes
the evolution of the structure functions by a power series in $1/Q^2$.
The first moment of $g_1$ takes the form~\cite{Ji:1993sv}:
\begin{equation}
\label{eq4.4}
\int_0^1dx\ g_1(x,Q^2) = \tilde{\Gamma}_1 +
(a_2+4d_2+4f_2)\,\frac{M^2}{9Q^2}
 + {\mathcal{O}}\,\left(\frac{M^4}{Q^4}\right)\ ,
\end{equation}
with only a logarithmic $Q^2$ dependence of $\tilde{\Gamma_1},\ a_2,\ d_2$, and
$f_2$. In the nomenclature of the OPE, the leading term
$\tilde{\Gamma}_1$ is twist 2, $a_2$ is a target mass correction and also of
twist 2, and $d_2$ and $f_2$ are matrix elements of twist-3 and twist-4 quark
gluon operators, respectively. Several model estimates as well as lattice QCD calculations have been performed
for the twist-3 matrix element $d_2$. The inelastic contribution
is given by~\cite{Kao:2003jd}:
\begin{equation}
\label{eq4.5}
d_2^{\rm{inel}}(Q^2) = \frac {Q^4}{8 M^4} \left\{ I_1(Q^2) - I_{TT}(Q^2) +
\frac{M^2Q^2}{\alpha_{em}} \, \delta_{LT}(Q^2) \right\} \, .
\end{equation}
Because $I_{TT}(0) = I_1(0) = -\kappa^2/4$, the RHS of this equation is
determined by the slopes of the generalized GDH integrals at the real photon
point, and therefore $d_2(Q^2)$ can be predicted also by ChPT, at least for
sufficiently small $Q^2$. The lower panel of Fig.~\ref{fig_Gamma2n_d2n} compares the recent data of the JLab E94-010
Collaboration~\cite{Amarian:2003jy} for $d_2^n(Q^2)$ to the predictions of MAID and ChPT~\cite{Kao:2002cp,Bernard:2002pw}.
\section{Conclusions}
Since the 1980s, deep inelastic scattering (DIS) at CERN, HERMES, and SLAC has provided invaluable information on the
nucleon structure functions and their moments. In particular, the Bjorken sum rule, a strict prediction of
QCD, was shown to agree with the data. However, a complete picture of the spin dynamics requires
measurements over the full range of momentum transfer, from long-range (coherent) to short-range (incoherent)
phenomena. For this reason, several dedicated experiments were launched to study the nucleon's spin structure
from photoproduction to electroproduction at low and intermediate momentum transfer. As a result of these efforts,
a wealth of new precision data has been assembled over the past decade, and more data are expected
in the coming years.
\newline \indent
Recent photoproduction experiments at MAMI and ELSA have proved the saturation of the Gerasimov-Drell-Hearn (GDH)
sum rule for the proton at photon energies of $2-3$~GeV, that is, the physics of
the anomalous magnetic moment of the nucleon is essentially determined by the resonance region. However, the Regge
tail of the integrand may provide a contribution of $5-10~\%$, and therefore an experiment at higher energies
would be helpful. Because nuclear and nucleonic degrees of freedom are intertwined, the situation for the
neutron deserves further studies, such as the comparison of results derived from different ``neutron targets''
and measurements of all the decay channels (incoherent and coherent pion production, non-pionic decays).
Due to the energy-weighting, the (leading) spin polarizability is now well known. However, the experiments have also
the potential to determine the higher polarizabilities and, through DRs, the full amplitude
for forward scattering.
\newline \indent
Collaborations in Halls A, B, and C of the JLab have provided a rich body of high-quality data with virtual
photons. The generalized GDH integral and related integrals were investigated from very small ($Q^2=0.01~{\rm {GeV}}^2$)
to quite large ($Q^2=5~{\rm {GeV}}^2$) momentum transfer and over the full resonance region ($W\leq 3~{\rm {GeV}}$).
The data for the generalized GDH integral of the proton are characterized by a rapid variation with $Q^2$ from
large negative values for real photons to a zero at $Q^2\approx 0.2~{\rm {GeV}}^2$, followed by positive values in the $1/Q^2$ tail
towards the Bjorken scaling region. Quite spectacular is the new information on the second (longitudinal-transverse) spin
structure function. The neutron data support the validity of the Burkhardt-Cottingham (BC) sum rule, a fascinating prediction connecting
an integral over the excitation spectrum to ground state properties (nucleon form factors) at all values of momentum
transfer, that is, at all distances. At $Q^2=1.28~{\rm {GeV}}^2$, a recent experiment confirmed the BC sum rule also for the proton.
Systematic investigations over the full range of momentum transfer are prerequisite to prove the sum rule beyond doubt.
As for real photons, the neutron results should be checked by extraction from different nuclear targets, in particular with regard
to nuclear corrections at small momentum transfer.
\newline \indent
The recent JLab experiments have also determined the two generalized spin polarizabilities of the nucleon, which yield
information on the spatial distribution of the polarization densities. The rapid decrease with momentum transfer indicates
that these polarizabilities are completely dominated by long-range phenomena, an interplay of non-resonant pion production
near threshold and resonance excitation. Contrary to earlier expectations, ChPT has not been very successful in describing
the fine tuning of these two contributions at small values of $Q^2$. In the scaling region ($Q^2> 4-5~{\rm {GeV}}^2$),
the DIS contributions to the polarizabilities become important. However, in this region the polarizability has dropped by $3-4$ orders
of magnitude relative to the real photon point. As for real photons, experiments with virtual photons have
the potential to determine the higher generalized polarizabilities and thus the full forward amplitude.
The Hall A Collaboration has also investigated the evolution of higher-twist functions to low $Q^2$. The continuation of
the dynamic twist-3 moment $d_2$ peaks at $Q^2\approx 0.5~{\rm {GeV}}^2$ and can be well described by resonance contributions.
The strong variation with $Q^2$ indicates the importance of higher terms if the twist expansion is extended towards
$Q^2\approx 1~{\rm {GeV}}^2$. A continuation of these experiments to larger $Q^2$ is of high interest in order to
determine the point at which $d_2$ approaches its asymptotic value.
\newline \indent
In summary, considerable progress has been made in our qualitative
understanding of the nucleon's spin structure over the full range of momentum transfer.
As the virtuality of the photon increases, the strongly correlated
many-body system ``nucleon'' is seen through a microscope with better and better
resolution, and the pertinent degrees of freedom change from Goldstone bosons
and collective resonances to the primary constituents, the quarks and
gluons. Although we have witnessed a steady improvement of ab-initio calculations such as
ChPT and lattice QCD, a quantitative understanding of the inclusive processes in the resonance region is still
missing. However, the new inclusive experiments have proved without doubt that the resonance region of $W<2$~GeV plays
an important role over a wide range of momentum transfer. In order to fully comprehend the nature of
these resonances, semi-inclusive experiments are asked for, such as one-pion electroproduction in the
higher resonance regions. In particular, the considerable longitudinal
strength shown by the inclusive data above $W<1.5$~GeV may have important consequences for the
multipole analysis of one-pion resonance decay.
\newline \indent
We conclude that the wealth of new data has been both informative and challenging for
theory, and we look forward to further experimental and theoretical advances in
our quest to understand the nucleon's spin structure in the realm of non-perturbative QCD.
\begin{theacknowledgments}
The author is grateful to Annual Reviews and Elsevier for permissions to use figures from
previous publications. Special thanks go to J.P. Chen,
S.E. Kuhn, O.A. Rond\'{o}n, and K. Slifer. This work was supported by Deutsche Forschungsgemeinschaft (SFB 443).
\end{theacknowledgments}
\bibliography{drechsel_SpinStructure}
\end{document}